\documentclass[pre,aps,floatfix,showpacs,preprint]{revtex4}
\usepackage{graphicx}
\usepackage{epsfig,graphics}
\usepackage{amsmath, amsthm, amssymb}

\newcommand{\be}{\begin{equation}}
\newcommand{\ee}{\end{equation}}
\newcommand{\bea}{\begin{eqnarray}}
\newcommand{\eea}{\end{eqnarray}}

\begin{document}
\title{Regular and Chaotic States in a Local Map Description of
Sheared Nematic Liquid Crystals}
\author{S. M. Kamil}
\email{kamil@imsc.res.in} 
\affiliation{The Institute of Mathematical Sciences, 
C.I.T. Campus, Taramani, Chennai 600013, India}
\author{Sudeshna Sinha}
\email{sudeshna@imsc.res.in}
\affiliation{The Institute of Mathematical Sciences, 
C.I.T. Campus, Taramani, Chennai 600013, India}
\author{Gautam I. Menon}
\email{menon@imsc.res.in} 
\affiliation{The Institute of Mathematical Sciences, 
C.I.T. Campus, Taramani, Chennai 600013, India}
\date{\today}

\begin{abstract}
We propose and study a local map capable of describing the
full variety of dynamical states, ranging from regular
to chaotic, obtained when a nematic liquid crystal is
subjected to a steady shear flow. The map is formulated
in terms of a quaternion parametrization of rotations
of the local frame described by the axes of the nematic
director, subdirector and the joint normal to these,
with two additional scalars describing the strength of
ordering. Our model yields kayaking, wagging, tumbling,
aligned and coexistence states, in agreement with previous
formulations based on coupled ordinary differential
equations.  Such a map can serve as a building block for
the construction of lattice models of the complex
spatio-temporal states predicted for sheared nematics.
\end{abstract}
\date{\today}
\maketitle

Driven complex fluids exhibit an unusual variety of
dynamical states\cite{Diat,schmitt,berret0,
eiser, sood, sood1, salmon}.  When such fluids are sheared
uniformly, the stress response is regular at very small
shear rates.  However, at larger shear rates the response
is often intrinsically unsteady, exhibiting oscillations
in space and time as a prelude to intermittency and
chaos\cite{sood,sood1,Cladis}.  Such chaos associated
with rheological response or ``rheochaos'', occurs in
regimes where the Reynolds number is very small. It must
thus be a consequence of constitutive and not convective
non-linearities, originating in the coupling of the flow to
structural or orientational variables describing the local
state of the fluid\cite{Cates_Fielding, Olmsted_Faraday}.
The diverse possibilities for internal degrees of freedom
in complex fluids, such as the orientation of nematogenic
molecules, layer stacking in lamellar and onion phases and
heterogeneities arising from local jamming in colloidal
suspensions, implies that the study of the rheology of
complex fluids should illuminate a variety of non-trivial
steady states in driven soft matter.

Recent rheological studies of  ``living polymers'',
solutions of worm-like micelles in which the energies
for scission and recombination are thermally accessible,
obtain an oscillatory response to steady shear at low
shear rates which
turns chaotic at larger shear rates\cite{sood,sood1}. It
has been argued that a hydrodynamic description of
this behaviour requires a field describing the local
orientation of the polymer, motivating a 
treatment of the problem of an orientable fluid,
such as a nematic, in a uniform shear flow\cite{hess30,doi,MD}.
Nonlinear relaxation equations for the symmetric,
traceless second rank tensor ${\bf Q}$ characterizing
local order in a sheared nematic have been derived
\cite{hess30,doi,MD,hess36,hess20,MD1,Olmsted,HS}. 
Assuming spatial uniformity, a system of 5
coupled ordinary differential equations (ODEs) for the 5
independent components of ${\bf Q}$ in a suitable tensor
basis is obtained. Solving this system of equations yields
a complex phase diagram admitting many states -- aligned,
tumbling, wagging, kayak-wagging, kayak-tumbling and
chaotic -- as functions of the shear rate $\dot{\gamma}$
and a phenomenological relaxation time which is a parameter
in the equations of motion\cite{GR1,GR2,Grosso}. Recent work adds
spatial variations: numerical studies of the partial
differential equations thus obtained yield a phase diagram
containing spatio-temporally regular, intermittent and
chaotic states\cite{sr,sr1}.

The degrees of freedom which enter a coarse-grained
description of an orientable fluid are mesoscopic.
Spatio-temporal structure arises from the coupling
of locally ordered regions, through processes such
as molecular diffusion, flow-induced dissipation
and advection.  A powerful approach to understanding
complex spatio-temporal dynamics is based on the study
of coupled map lattices, a numerical scheme in which
maps placed on the sites of a lattice evolve both
via local dynamics as well as through couplings to
neighbouring sites\cite{kaneko}. However, the utility of
this methodology in a specific context is often severely
limited by the availability of local maps able to describe
the spatially uniform case.  This paper addresses this
requirement in the context of a model for rheochaos,
proposing the first local map description of the regular
and chaotic states obtained in sheared nematics.

There is, in general, no systematic procedure for the
construction of such maps. However, it is reasonable to
require that any such map should accurately reproduce
the full variety of states obtained through the study
of the corresponding ODEs. It should also enable useful
physical insights through a sensible choice of physical
variables. One obvious possibility
is simply the discretization of the governing ODEs.
Such a choice of variables,
however, is not particularly illuminating as these
equations are formulated in terms of the components of
${\bf Q}$ in a specific space-fixed tensor basis, rather
than in terms of variables more natural to the problem.

\begin{figure}
\begin{center}
\includegraphics[width=3in]{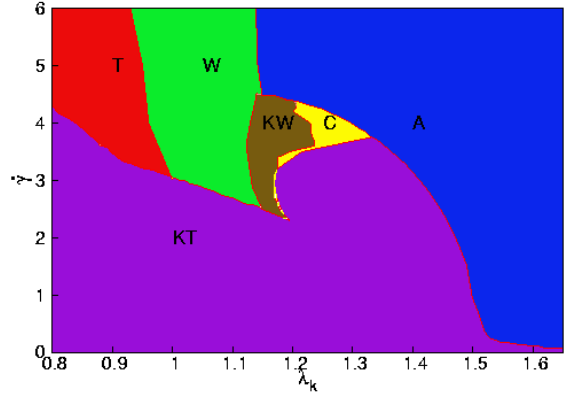}
\caption{[Color online] The phase diagram of steady states
in our model, illustrating regimes in which the following
steady states are obtained for a generic initial condition: an
aligned state denoted as `A', a tumbling state labelled as
`T', a wagging state `W', a kayak-tumbling state `KT',
a kayak-wagging state denoted by `KW' and a complex state
denoted as `C'. This phase diagram closely resembles phase
diagrams plotted in Refs.~\cite{GR2}.
}
\label{phasedia}
\end{center}
\end{figure}

We have thus explored an alternative formulation of this
problem, constructing a local map in terms of quaternion
variables. These variables encode the dynamics of the
orthogonal set of axes associated with
the eigenvectors of ${\bf Q}$, {\it i.e.} the 
director, sub-director and the joint normal to these. Our
approach incorporates biaxiality, is formulated in terms
of physically accessible variables and is computationally
straightforward to implement.  Our results,
summarized in the phase diagram of Fig.~\ref{phasedia},
are in good agreement with previous work based on ODEs
\cite{GR2}, but provide an
efficient alternative to such methods\cite{footnote2}.

Defining $\widehat{\bf b}: = \frac{1}{2}({\bf b} + {\bf b}^T) - 
\frac{1}{3}(tr{\bf b}) {\bf \delta}$
to be the symmetric-traceless part of the second-rank
tensor {\bf b}, the equation of motion for ${\bf Q}$ in a
passive velocity field is \cite{hess30,GR2}:
\begin{equation}\label{one}
\frac{d{\bf Q}}{dt} - 2\widehat{{\bf\Omega \cdot Q}} - 
2\sigma \widehat{{\bf \Gamma \cdot Q}} + 
\tau_Q^{-1}{\bf \Phi} = - \sqrt{2}\frac{\tau_{ap}}{\tau_a}{\bf\Gamma}
\end{equation}
where the tensor 
${\bf \Omega} = \frac{1}{2}((\nabla{\bf v})^T - \nabla{\bf v})$,
${\bf \Gamma} = \frac{1}{2}((\nabla{\bf v})^T + \nabla{\bf v})$
and $\nabla{\bf v}$ is the velocity gradient tensor, 
with ${\bf v} = \dot{\gamma}y{\bf e^x}$, where
${\bf e^x}$ is a unit vector in the $x-$ direction.
The velocity is along the $x$ direction,
the velocity gradient is along the $y$ direction, while
$z$ is the vorticity direction.
The quantities
$\tau_a > 0$ and $\tau_{ap}$ are phenomenological quantities
related to relaxation times,
$\sigma$ describes the change of alignment 
caused by ${\bf \Gamma}$ and ${\bf \Phi} = \partial \phi/\partial{\bf Q}$, with
$\phi({\bf Q}) = \frac{1}{2}A{\bf Q:Q} - 
\frac{1}{3}\sqrt{6}B({\bf Q \cdot Q}):{\bf Q} + \frac{1}{4}C({\bf Q:Q})^2$.
The notation $Q:Q$ represents $Q_{ij}Q_{ji}$, with repeated indices
summed over.
Here $A = A_0(1-T^*/T)$, and $B$ and $C$ are constrained by the 
conditions $A_0 > 0$, $B>0,C>0$ and $B^2 >\frac{9}{2}A_0C$. 
Scaling $t = t^*\tau_a/A_k$, ${\bf v}= {\bf v^*}A_k/\tau_a$ 
and $a = a^*a_k$, Eqn.~(\ref{one}) can be written in dimensionless form,
$\frac{d{\bf Q^*}}{dt^*} - 2\widehat{{\bf\Omega^*.Q^*}} - 2\sigma \widehat{{\bf \Gamma^*.Q^*}} + (\theta {\bf Q}^* - 3\sqrt{6}\widehat{{\bf Q^*.Q^*}}
	+ 2({\bf Q^*:Q^*)Q^*})
= \sqrt{\frac{3}{2}}\lambda_k{\bf\Gamma^*}$
where $A_k = A_0(1 - T^*/T_k)=2B^2/9C, a_k = a_{eq}(T_k)=2B/3C$ is 
the (nonzero) equilibrium value
of the scalar order parameter $a$ at the transition temperature $T_k$,\ \ $\lambda_k = -\frac{2}{3}\sqrt{3}\frac{\tau_{ap}}{\tau_a a_k}$\ and 
$\theta = (1 -\frac{T^*}{T})/(1 -\frac{T^*}{T_k})$ is the reduced temperature.
\begin{figure}
\begin{center}
\includegraphics{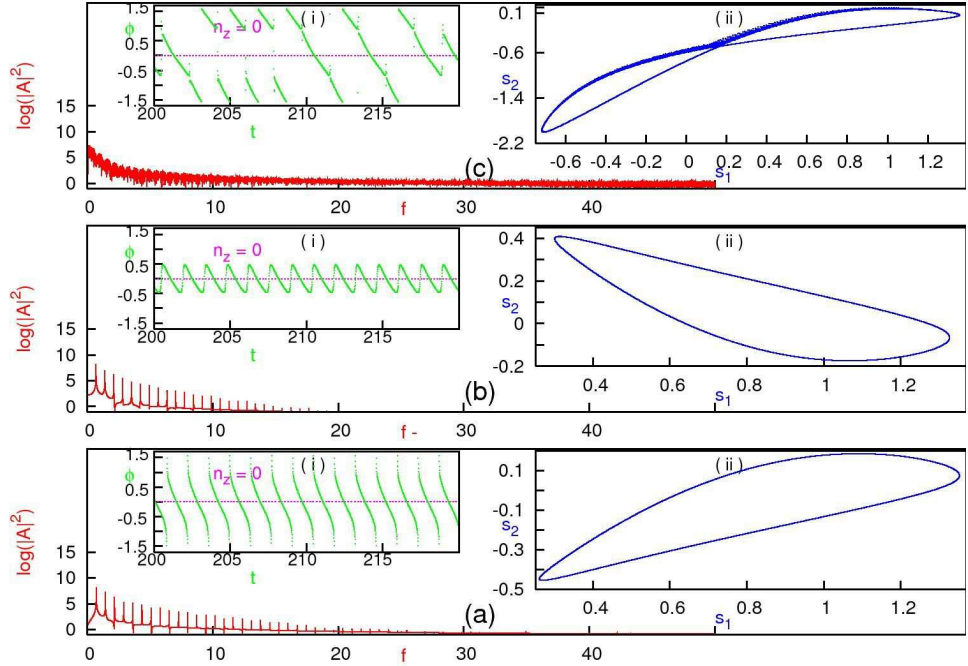}
\caption{[Color Online] The sequence of three main panels 
shows the power spectrum 
associated
with states in the regimes labelled (a) T and (b) W in the
phase diagram of Fig.~\ref{phasedia}. The topmost panel (c) shows
a mixed state (M) (not shown separately in Fig.~\ref{phasedia}),
associated with the boundary between W and T 
The inset labelled (i) in all these panels shows typical
plots of the time-dependence of the z-component of the director $n_z$
and the angle $\phi$ made by the projection of the director on the
$x-y$ plane with the $x-$ axis.
The insets labelled (ii) in all these panels show the trajectory
in the $s_1 -s_2$ plane.
}
\label{state1}
\end{center}
\end{figure}

The ${\bf Q}$ tensor admits the following parametrization:
$Q_{ij} = \frac{3s_1}{2} \left ( n_i n_j - \frac{1}{3}\delta_{ij} \right )
+ \frac{s_2}{2} \left ( m_i m_j - l_i l_j \right )$,
where $s_1$ and $s_2$ represent the magnitude of the
ordering along {\bf n} (the director) and {\bf m} (the
subdirector), with {\bf n} and {\bf m} unit vectors and
{\bf l} = {\bf n} $\times$ {\bf m}. The dynamics of ${\bf
Q}$ thus involves both the dynamics of the frame defined
by ${\bf n},{\bf m}$ and ${\bf l}$ as well as 
the dynamics of $s_1$ and $s_2$. The frame dynamics
can be represented in many equivalent ways, such as
through coordinate matrices, axis-angle or Euler angle
representations. However, the coordinate matrix representation 
requires a large number of parameters,
the axis-angle representation suffers from redundancy and
the use of the Euler-angle representation is marred by the
``gimbal-lock'' problem\cite{gimbal-lock}.
Our parametrization of the frame dynamics uses
quaternion variables, providing an elegant, compact and
numerically stable alternative to these representations.

Equations for $\dot{\bf{n}}$, $\dot{\bf{m}}$ and
$\dot{\bf{l}}$ as well as for the order parameter
amplitudes $\dot{s_1}$ and $\dot{s_2}$ can be derived by
considering a reference frame in which the director and
subdirector are stationary (body frame). In the body frame,
denoted by primed vectors, the director can be chosen
to be ${\bf n}^\prime = (1,0,0)$, the subdirector to
be ${\bf m}^\prime = (0,1,0)$, with ${\bf l}^\prime =
(0,0,1)$.  The transformation matrix ${\bf A}$
which maps vectors from the lab frame to the body frame,
can be defined in terms of quaternion
parameters $(e_0,\ldots,e_4)$ constrained by
$e_{0}^{2} + e_{1}^{2} + e_{2}^{2} + e_{3}^{2} = 1$\cite{amatrix}.  
The quantities ${\bf n} = (n_x,n_y,n_z), {\bf m} = 
(m_x,m_y,m_z)$ and ${\bf l} = (l_x,l_y,l_z)$ are
easily obtained using this mapping,
yielding ODE's for the parameters 
${s_1,s_2,e_0,e_1,e_2,e_3,e_4}$. These are converted
into a map using a first-order Euler scheme.
After each discrete time step, we renormalise the quaternion
variable. Choosing $\sigma$ and $\theta$ equal to zero 
for all the results reported here in common with earlier
work, our map is then defined through
\bea {s_1}^{t+1} &= {s_1}^t + &\Delta \left(\frac{1}{6}\ \{9\
\sqrt{6}\ s_1^2 - 18\ s_1^3 - 3\ \sqrt{6}\ s_2^2 -
	6\ s_1\ s_2^2 +
				3\ \sqrt{6}\ n_{x}\
				n_{y}\ \dot{\gamma}\
				\lambda_{k}\}\right)^t
		\nonumber \\
		{s_2}^{t+1} &= {s_2}^t + &\Delta \left( -3\ \sqrt{6}\ s_1\ s_2 - 3\ s_1^2\ s_2 -
    s_2^3 - \sqrt{\frac{3}{2}}\
    (l_{x}\ l_{y} -
	  m_{x}\ m_{y})\ \dot{\gamma}\ \lambda_{k}\right)^t
	  \nonumber \\
		{e_0}^{t+1} & = {e_0}^t + &
\Delta \Bigg(\frac{1}{4}\dot{\gamma}\ e_{3} +
    \frac{1}{4}\ \sqrt{\frac{3}{2}}\ \dot{\gamma}\
    \{-\frac{(l_{y}\ m_{x} +
		      l_{x}\ m_{y})\ e_{1}}{s_2}
		      + \frac{2\ (l_{y}\
		      n_{x} + l_{x}\ n_{y})\ 
e_{2}}{3\ s_1 + s_2}\nonumber\\& & + \frac{2\ (m_{y}\
n_{x} + m_{x}\ n_{y})\ e_{3}}{-3\ s_1 + \ s_2}\}\
\lambda_{k}\Bigg)^t \nonumber\\
{e_1}^{t+1}& = {e_1}^t + &\Delta \Bigg( \frac{1}{4}\
\dot{\gamma}\ e_{2}\  +
    \frac{1}{4}\ \sqrt{\frac{3}{2}}\ \dot{\gamma}\
    (\frac{(l_{y}\ m_{x} + l_{x}\ m_{y})\ e_{0}}{s_2}
    -\frac{2\ \
(m_{y}\ n_{x} + m_{x}\ n_{y})\ e_{2}}{-3\ s_1 +
s_2}\nonumber \\ & & + \frac{2\ (l_{y}\ n_{x} +
l_{x}\ n_{y})\ \ e_{3}}{3\ s_1 + s_2})\ \lambda_{k}\Bigg)^t
\nonumber \\
{e_2}^{t+1} & = e_2^t + &
\Delta \Bigg(-\frac{1}{4}\ \dot{\gamma}\ e_{1}\  +  \frac{1}{4}\
\sqrt{\frac{3}{2}}\ \dot{\gamma}\ (-\frac{2\ (l_{y}\
n_{x} +
		      l_{x}\ n_{y})\ e_{0}}{3\ s_1 +
		  s_2} + \frac{2\ (m_{y}\ n_{x} +
		  m_{x}\ n_{y})\ e_{1}}{-3\ s_1 \
+ s_2}\nonumber\\& & + \frac{(l_{y}\ m_{x} + l_{x}\
m_{y})\ e_{3}}{s_2})\ \lambda_{k}\Bigg)^t
		\nonumber \\
		{e_3}^{t+1}& = e_3^t + &
\Delta \Bigg(-\frac{1}{4}\ \dot{\gamma}\ e_{0}\   +
    \frac{1}{4}\ \sqrt{\frac{3}{2}}\ \dot{\gamma}\
    (-\frac{2\ (m_{y}\ n_{x} +
		      m_{x}\ n_{y})\ e_{0}}{-3\ s_1 +
		  s_2} -\frac{ 2\ (l_{y}\ n_{x} +
		  l_{x}\ n_{y})\ e_{1}}{3\ s_1 + s_2}\
 \nonumber\\& &- \frac{(l_{y}\ m_{x} + l_{x}\ m_{y})\
 e_{2}}{s_2})\ \lambda_{k}\Bigg)^t
\end{eqnarray}
We choose $\Delta = 0.01$ 
for all our calculations. (The phase boundaries shown in 
Fig.~\ref{phasedia} exhibit a weak 
dependence on $\Delta t$. However, provided $\Delta t$ is 
chosen small enough, 
this dependence may be neglected.)
The superscript `t' indicates that the
values of the variables are taken at the t'th discrete time step.
The control parameters are the dimensionless shear rate
$\dot{\gamma}$ and $\lambda_{k}$.
In place of the 5 coupled ODE's used in the conventional
parametrization of the dynamics of ${\bf Q}$, we have 6
equations constrained by the normalization requirement, thereby
conserving the number of degrees of freedom.

\begin{figure}
\begin{center}
\includegraphics{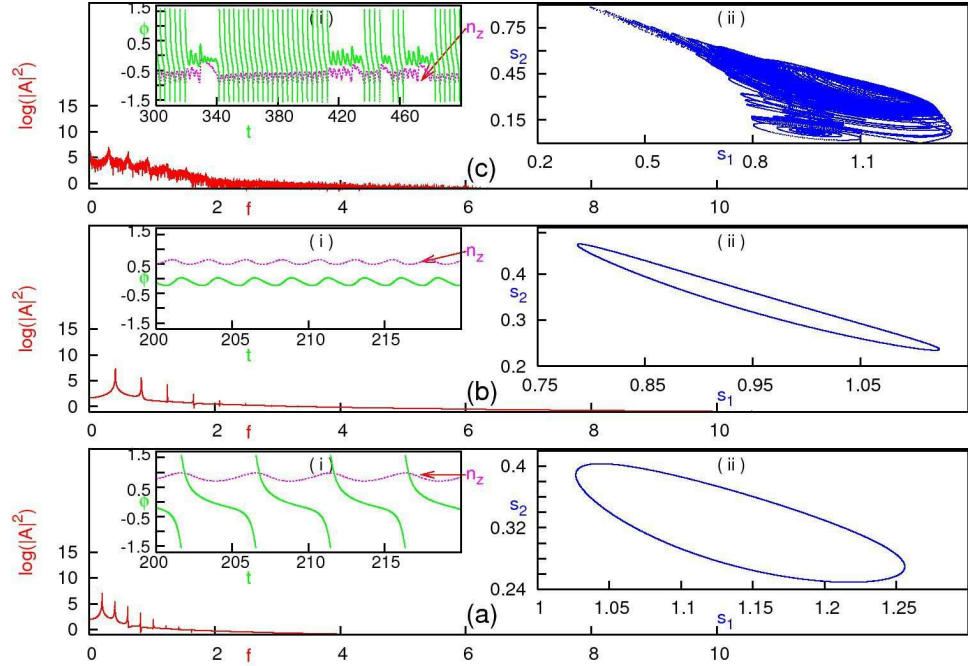}
\caption{[Color online] The sequence of three main panels shows the power spectrum associated
with states in the regimes labelled (a) KT (kayak-tumbling), 
(b) KW (kayak-wagging) and (c) C (complex or chaotic) in the
phase diagram of Fig.~\ref{phasedia}. 
The inset labelled (i) in all these panels shows typical
plots of the time-dependence of the z-component of the director $n_z$
and the angle $\phi$ made by the projection of the director on the
$x-y$ plane with the $x-$ axis.
The insets labelled (ii) in all these panels show the trajectory
in the $s_1 - s_2$ plane.
}
\label{state2}
\end{center}
\end{figure}

In our numerical analysis of the map, we start
typically from random initial conditions, omitting
sufficient transients ($\sim 10^5$ time steps) to
ensure that the asymptotic attractor of the dynamics
is reached. Our analysis includes inspection of the (i)
power spectrum, (ii) phase portraits, (iii) bifurcation
diagrams and  (iv) time series of the different relevant
variables. 
Figs. ~\ref{state1} and ~\ref{state2} show the
variety of states obtained in our numerical calculations.
Each sub-figure, labelled as
Figs.~\ref{state1} (a) - (c) and Figs. ~\ref{state2}
(a)-(c), has the following structure: The first inset,
labelled (i) for all figures, describes the time dependence
of $n_z$, the z-component of the director, and the angle $\phi$
made by the projection of the director on the $x-y$
plane with the $x-$ axis.  The second inset, labelled
(ii) for all figures, plots the quantities measuring the
amount of ordering along director and sub-director against
each other, providing the attractor of the system in the
$s_1-s_2$ plane for a generic initial condition. The main
panel in each of the sub-figures shows the power spectrum
of $s_1$, $\ln (|A(f)|^2)$ against frequency $f$ on a semi-log plot.

The following states are easily identified:
(I) An {\bf Aligned} state denoted as `A' in the phase-diagram of
Fig.~\ref{phasedia}, but omitted, for brevity, from the states
shown in Fig.~\ref{state1}  and Fig.~\ref{state2}.  In the aligned
state, neither the frame orientation,
nor $s_1$ and $s_2$, vary in time.  The director is
aligned with the flow at a fixed angle;
(II) A {\bf Tumbling} state, in which the director lies in the shear
plane (the $xy$ plane) and rotates about the vorticity
direction (the $z$ axis). Fig.~\ref{state1}(a)(i)
indicates that this state is a stable in-plane state, since the
z-component of the director is zero. Also, the angle made by the
projection of the director on the x-y plane varies smoothly between
$\pi/2$ and -$\pi/2$.  Fig.~\ref{state1}(a)(ii) shows the
periodic character of this state.  This state is labelled as `T' in
the phase-diagram of Fig.\ref{phasedia};
(III) A {\bf Wagging} state, in which the director lies in the shear
plane, but oscillates between two values.  Note that
Fig. ~\ref{state1} (b)(i) indicates that this state is a stable
in-plane state.  Also, the director oscillates back and forth
in-plane as indicated in Fig. ~\ref{state1} (b)(ii).
Fig. ~\ref{state1} (b) shows that this state is a periodic state
with sharp delta-function peaks in the power spectrum.  These states
are denoted as `W' in the phase-diagram in Fig.\ref{phasedia}. In
addition, we obtain 
(IV) A {\bf Kayak-Tumbling} state, equivalent to the tumbling state,
but in which the director is out of the shear plane.  Thus, as shown
in Fig.~\ref{state2}(a) $n_z \neq 0$ and the projection of the
director on the $xy$ plane rotates through a full cycle. Such states
are temporally periodic, as shown in Fig.~\ref{state2}(a); the
regular cycles evident in the map of $s_1$ vs. $s_2$ 
(Fig.~\ref{state2}(a)(ii)) is a further
indication of periodic behaviour. These states are noted as `KT' in
the phase-diagram of Fig. ~\ref{phasedia};
(V) A {\bf Kayak-Wagging} state where, as in KT, the
director is out of plane, but the projection of the director on the
shear plane oscillates between two values. The properties of such
states are illustrated in Fig.~\ref{state2}(b).  Such states are
again temporally periodic. The cyclic trajectory of the system in
the $s_1-s_2$ plane (Fig.~\ref{state2}(b)(ii))
further confirms such periodic
behaviour. These states are denoted by `KW' in the phase-diagram of
Fig.~\ref{phasedia};
(VI) A {\bf Mixed} state, typically found close to the boundaries
between wagging and tumbling states, whose properties are
illustrated in Fig.~\ref{state1}(c).  In such states, the director
exhibits both oscillation and complete rotations. Power spectra
obtained at the boundaries of this phase, for example near
$\lambda_k = 0.99$ and $\dot{\gamma} = 4.0$, have a broad range of
frequencies, and, (VII) A {\bf Complex} state, in which the director 
lies out of the
shear plane but both oscillates and rotates. The complex phase
exhibits chaotic behaviour, as can be seen in
Fig.~\ref{state2}(c). Note that the delta function peaks in the
power spectrum exhibited by the periodic states discussed earlier
have broadened into a continuum of frequencies. The plot of $s_1$
vs. $s_2$ displays no regular structure. These state are noted as
`C' in the phase-diagram in Fig.\ref{phasedia}.
In addition to these states, we also obtain a log 
rolling state in which the director is perpendicular to the shear 
plane (not shown).

The range of dynamical states manifest in this problem
is also evident from bifurcation diagrams obtained at
fixed $\dot{\gamma} = 4.0$, varying $\lambda_k$.  Such a
cut intersects KT, T, W, KW, C and A states in the phase
diagram. The quantities $n_z$ and the Poincare section of
$s_1$ show that $n_z = 0$ for the T, W and A states,
while the KT, KW and C states are out-of-plane states with
$n_z \neq 0$.  Further, the $s_1$ section, displays a fixed
point for the aligned state, regular cycles for the KT,
T W and KW states and irregular (chaotic) behavior for
the C state.

Aradian and Cates have recently studied a minimal model
for rheochaos in shear-thickening fluids, using equations
which describe a shear-banding system coupled to a retarded
stress response\cite{aradian_cates}. These authors connect
their model system to a modified Fitzhugh-Nagumo model,
a dynamical system with a variety of interesting and
complex phases. Fielding and Olmsted study instabilities in
shear-thinning fluids, where the instability originates
in the multi-branched character of the constitutive
relation\cite{fielding_olmsted}.  Chakrabarty {\it et al.}
report a study of the PDE's describing the dynamics of
${\bf Q}$, characterizing spatio-temporal routes to chaotic
behaviour in sheared nematics~\cite{sr}.  These studies
allow for spatial variation - although restricted so
far to the one-dimensional case - whereas our local
map describes the spatially uniform situation. However,
the dynamical system we study is obtained directly from
the underlying dynamics, in contrast to the approaches
of Refs.~\cite{aradian_cates,fielding_olmsted}.
Whether coupling maps of the sort we construct permits a
complete description of the spatio-temporal structure
obtained in Ref.~\cite{sr} remains to be seen.

In conclusion, we have proposed a local map
describing the variety of dynamical states obtained in
a model for sheared nematics. Our phase diagram,
Fig.~\ref{phasedia},
contains all non-trivial dynamical states obtained in
previous work. It also closely resembles, even
quantitatively, phase diagrams obtained in
previous work which used ordinary differential equations
formulated in continuous time.  Our work thus supplies a
crucial ingredient in the construction of coupled
map lattice approaches to the spatio-temporal aspects of
rheological chaos, a problem currently at the boundaries
of our understanding of the dynamics of complex fluids.

\acknowledgments The authors thank A.K. Sood, Sriram
Ramaswamy, Chandan Dasgupta and Ronojoy
Adhikari for discussions. This work was partially supported 
by the DST (India) (GIM).


\begin{thebibliography}{}
\bibitem{footnote2} As is common in high dimensional complex
systems, there is a possibility of coexistence of
different dynamical states; our phase diagram shows the dominant
attractor of the dynamics.
\bibitem{minor} In contrast
to Ref.~\cite{GR2}, the phase boundary between $T$
and $KT$ phases is somewhat diffuse, admitting the
possibility of coexistence states sufficiently close to the
boundary. Such states become increasingly rare
at large values of $\dot{\gamma}$ and are suppressed as 
$\Delta$ is made smaller. Also, the chaotic region
has a marginally different shape in our calculations.
\end{thebibliography}
\end{document}